\documentclass{article}

\PassOptionsToPackage{numbers,sort&compress}{natbib}
\usepackage[preprint]{style_ZJ}

\usepackage[utf8]{inputenc} 
\usepackage[T1]{fontenc}    
\usepackage{hyperref}       
\usepackage{url}            
\usepackage{booktabs}       
\usepackage{amsfonts}       
\usepackage{nicefrac}       
\usepackage{microtype}      
\usepackage{amsmath,amsthm,amsfonts,amssymb,amscd}
\usepackage{mathtools}
\usepackage{lastpage}
\usepackage{enumerate}
\usepackage{fancyhdr}
\usepackage{mathrsfs}
\usepackage[x11names]{xcolor}
\usepackage{graphicx}
\usepackage{listings}
\usepackage{subcaption}
\usepackage[english]{babel}
\usepackage{apxproof}
\usepackage{thmtools}
\usepackage{thm-restate}
\usepackage{enumitem}
\usepackage{float}
\usepackage{bm}
\usepackage{wrapfig}
\usepackage{algorithm}
\usepackage{algpseudocode}

\usepackage{sidecap}
\usepackage{siunitx}
\newtheorem{theorem}{Theorem}
\newtheorem{corollary}{Corollary}[theorem]

\theoremstyle{definition}
\newtheorem{definition}{Definition}[section]
\newtheorem{example}[definition]{Example}

\definecolor{myGreen}{rgb}{0,0.5,0}

\DeclareMathOperator{\argmin}{argmin}

\DeclareMathOperator{\floor}{floor}

\title{A fast algorithm for All-Pairs-Shortest-Paths suitable for neural networks}

\author{
  Zeyu Jing and Markus Meister\\
  Division of Biology and Biological Engineering\\
  California Institute of Technology\\
  \texttt{\{zjing,meister\}@caltech.edu}
}

\begin{document}

\maketitle

\begin{abstract}
Given a directed graph of nodes and edges connecting them, a common problem is to find the shortest path between any two nodes. Here we show that the shortest path distances can be found by a simple matrix inversion: If the edges are given by the adjacency matrix $A_{ij}$ then with a suitably small value of $\gamma$ the shortest path distances are 

$$ D_{ij} = \operatorname{ceil} \left( \log_{\gamma} {\left[ {\left({\mathbf{I}}-\gamma {\mathbf{A}}\right)^{-1}} \right]}_{ij} \right)$$

We derive several graph-theoretic bounds on the value of $\gamma$, and explore its useful range with numerics on different graph types. Even when the distance function is not globally accurate across the entire graph, it still works locally to instruct pursuit of the shortest path. In this mode, it also extends to weighted graphs with positive edge weights. For a wide range of dense graphs this distance function is computationally faster than the best available alternative. Finally we show that this method leads naturally to a neural network solution of the all-pairs-shortest-path problem.

\end{abstract}

\section{Introduction} \label{sec:intro}

Many problems in animal behavior or in robotic control can be reduced to search on a graph. The graph may represent a spatial environment, like a road map, or a network of choices to be made in a cognitive task, like a game. Finding the shortest path from an initial state on the graph to a goal state is a central problem of graph theory \cite{zwickExactAndApproximate2001,lewisAlgorithmsFindingShortest2020}. Generally an ``all pairs shortest path'' (APSP) algorithm delivers a matrix containing the distance for all pairs of nodes on the graph. That matrix can then be used iteratively to construct the actual sequence of nodes corresponding to the shortest path. 

Much mathematical effort has focused on efficient ways to compute the pair-wise distance matrix starting from the matrix of adjacencies between nodes~\citep{chanMoreAlgorithmsAllPairs2010}.
The Floyd-Warshall algorithm \cite{floydAlgorithm97Shortest1962} is remarkably simple, consisting of three nested loops of conditional additions performed on an array.
However, for analog circuits, like the brain, this is an implausible solution to the problem. 
Here we present an alternative algorithm that computes distances on a graph using analog computation of the type performed by networks of neurons. 
If the graph adjacencies get stored in synapses of the network, the shortest distances emerge from the neural activities. 

We begin by introducing the proposed neural network and the function it computes, which we call the ``R-distance''. Then we derive a number of graph-theoretic bounds giving sufficient and necessary conditions for the R-distance to reflect the shortest path distances on the graph. We will show that the R-distance is an efficient APSP algorithm even on digital computers. Finally we return to practical constraints encountered in analog computing systems.

\section{An analog circuit for APSP} \label{sec:network}

\begin{figure}
  \centering
    \includegraphics[width=0.6\linewidth]{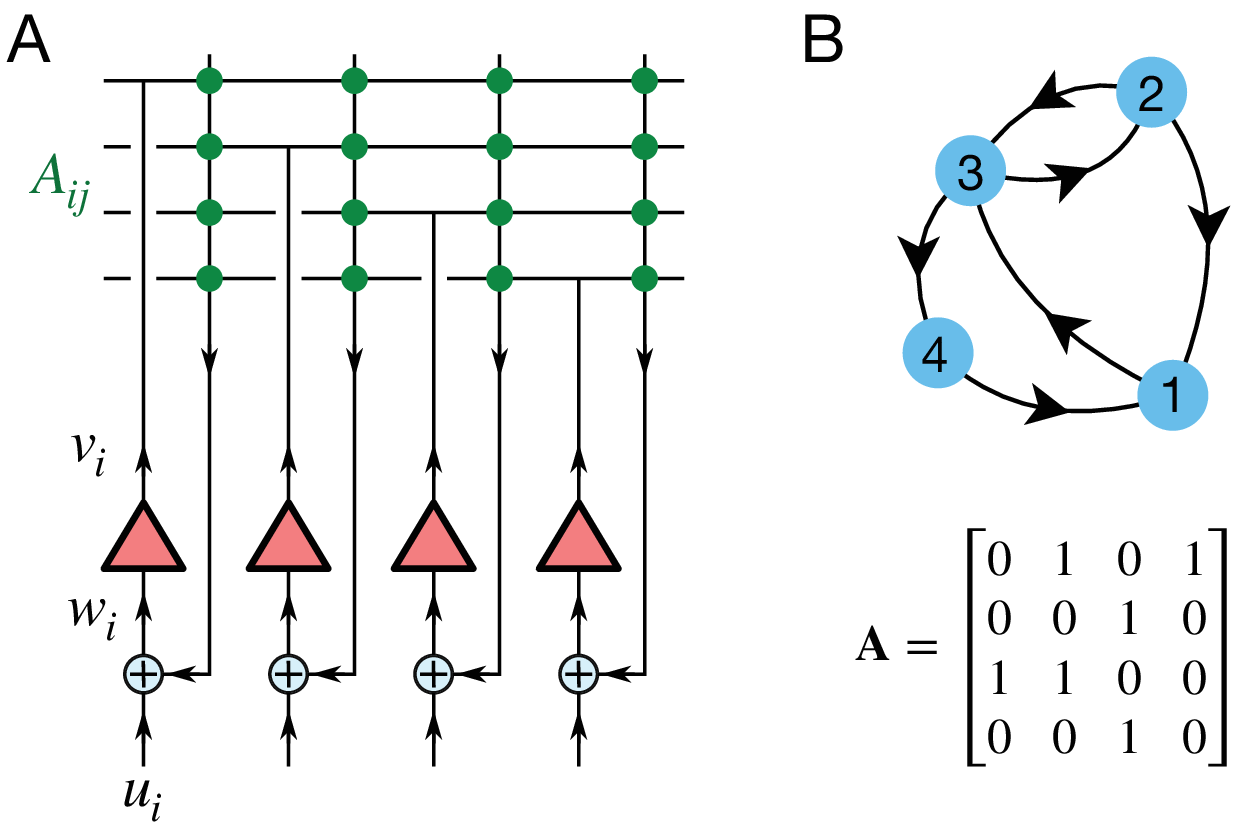}
    \caption{{\bf{Analog circuit to compute R-distances.}} {\bf{A:}} Each of the linear units (triangles) receives input $w_i$ and generates output $v_i=\gamma w_i$. The input is combined from an external drive $u_i$ and feedback from the outputs through the connections $A_{ij}$. {\bf{B:}} Example of a directed unweighted graph with 4 nodes and the corresponding adjacency matrix $A_{ij}$.}
\label{fig:circuit}
\end{figure}

Figure~\ref{fig:circuit}A shows a simple network of linear analog units with recurrent feedback. Each unit $i$ has an input $w_i$ that it converts to an output $v_i$:
\begin{equation}
v_i=\gamma \, w_i, \label{eq:network-output}
\end{equation}
where $\gamma$ is the gain of the units. The input consists of an external signal $u_i$ summed with a recurrent feedback through a connection matrix ${\mathbf{A}}$:
\begin{equation}
w_i=u_i + \sum_{ij} A_{ij} v_j \label{eq:network-input}.
\end{equation}

Combining Eqns~\ref{eq:network-output} and \ref{eq:network-input}, one finds the solution
\begin{equation}
{\vec v} = \gamma {\mathbf{Y}} {\vec u} ,  \label{eq:network-solution}
\end{equation}
where
\begin{equation}
{\mathbf{Y}} = \left(  {\mathbf{I}} - \gamma {\mathbf{A}} \right) ^ {-1} .  \label{eq:Y-defined}
\end{equation}

Suppose now that the connection matrix ${\mathbf{A}}$ represents the adjacency matrix of a directed unweighted graph (Fig~\ref{fig:circuit}B), namely
\footnote{Note $\mathbf{A}$ is the transpose of the usual convention for the adjacency matrix in graph theory. This definition is favored by the form of Eqn~\ref{eq:network-input} and we follow it throughout the text.} 
\begin{equation} \label{eq:adjacency}
{A_{ij}} = \left\{\begin{array}{*{20}{l}}
  {1, \; {\text{if there is a directed edge from node $j$ to node $i$}}} \\ 
  {0, \; {\text{otherwise}}} 
\end{array} \right. .
\end{equation}

We claim that one can set the gain $\gamma$ such that the network's transformation matrix $Y_{ij}$ is monotonically related to the matrix of shortest distances on the graph. Specifically, if we define 
\begin{equation} \label{eq:R-defined}
R_{ij}(\gamma)  = \log_{\gamma} Y_{ij}(\gamma) = \log_{\gamma} \left[ \left(  {\mathbf{I}} - \gamma {\mathbf{A}} \right) ^ {-1} \right]_{ij} ,
\end{equation}
then
\begin{equation} \label{eq:R-limit}
R_{ij}(\gamma) \xrightarrow{\gamma \rightarrow 0} D_{ij},
\end{equation}
where
\begin{equation} \label{eq:D-defined}
D_{ij} = \textrm{shortest path distance from node } j \textrm{ to node } i.
\end{equation}

Because $R_{ij}(\gamma)$ is closely related to the \emph{resolvent function} ${\left(\tfrac{1}{\gamma} {\mathbf{I}}- {\mathbf{A}}\right)^{-1}}$, we will call $R_{ij}(\gamma)$ the \textbf{R-distance} from node $j$ to $i$.

In the following sections we will prove the correspondence (\ref{eq:R-limit}) between the R-distance and the graph distance. For any practical computations one needs to use a finite gain $\gamma$ and we will find necessary and sufficient conditions for that value. We further generalize the results to weighted graphs, and illustrate use of the R-distance on diverse graph types.

\section{The asymptotic limit of small gain \texorpdfstring{$\gamma$}{}} \label{sec:asymptotic}

First, we show that the R-distance function of Eqn~\ref{eq:R-defined} delivers the correct graph distances in the asymptotic limit of $\gamma \rightarrow 0$ (Eqn~\ref{eq:R-limit}).
We start by generalizing a proof based on \citet{craneGeodesicsHeatNew2013}. 
This will set the stage for subsequent arguments extending the validity into the practical regime of nonzero $\gamma$.

\subsection{Unweighted directed graphs} \label{subsec:unweighted-directed-graph}

Given an unweighted directed graph 
\footnote{Unless specifically noted, all graphs are assumed strongly connected, i.e., for any pair of vertices $i,j$, $i \ne j$, there exists a directed path from $i$ to $j$ and a directed path from $j$ to $i$. Throughout this manuscript, graphs do not contain any loops or multiple edges.} 
with adjacency matrix $\mathbf{A}$ (Eqn~\ref{eq:adjacency}), 
consider the function 
\begin{align}
{\mathbf{Y}}(\gamma) &= \left({\mathbf{I}}-\gamma{\mathbf{A}}\right)^{-1} \label{eq:inverse} \\
&= {\mathbf{I}} + \gamma {\mathbf{A}} + \gamma^2 {\mathbf{A}}^2 + \dots \label{eq:inverse-series} .
\end{align}

The Taylor series in Eqn.~\ref{eq:inverse-series} converges if the gain $\gamma$ satisfies
\begin{equation}
0 \le \gamma < \frac{1}{\rho(\mathbf{A})},
\end{equation}
where $\rho(\mathbf{A})$ is the spectral radius of $\mathbf{A}$.

It is well-known that the 
powers of the adjacency matrix represent the effects of taking multiple steps on the graph~\citep{biggsAlgebraicGraphTheory1993},
$$
{\left[ {{{\mathbf{A}}^k}} \right]_{ij}} = N_{ij}^{(k)} = {\text{number of distinct walks to get from }}{j}{\text{ to }}{i}{\text{ in }}k{\text{ steps }}.
$$
so that
\begin{equation}
Y_{ij}(\gamma) = \sum_{k=0}^{\infty} N_{ij}^{(k)} \gamma^k . \label{eq:Y-series}
\end{equation}

Now, the shortest path distance from node $j$ to node $i$ is equal to the smallest nonnegative integer $k$ with a non-zero walk count $N_{ij}^{(k)}$:
$$
D_{ij} = \textrm{graph distance from } j \textrm{ to } i = \min\{k: N_{ij}^{(k)}>0\}.
$$

We define
$$
S_{ij} = N_{ij}^{(D_{ij})} = \textrm{number of distinct shortest paths from } j \textrm{ to } i.
$$

Then Eqn.~\ref{eq:Y-series} becomes
\begin{align}
Y_{ij}(\gamma) &= S_{ij} \gamma^{D_{ij}} + \sum_{k=D_{ij}+1}^{\infty} N_{ij}^{(k)} \gamma^k . \label{eq:Y-series-trunc}\\
&= \gamma^{D_{ij}} \left[ S_{ij} + C(\gamma) \right].
\end{align}

The sum of higher order terms $C(\gamma)$ can be bounded by a geometric series: The number of walks of length $k$ from node $j$ to $i$ cannot exceed the total number of walks of length $(k-1)$ starting from vertex $j$ and ending at any vertex. That number in turn cannot exceed $\Delta^{k-1}$, where $\Delta$ is the largest out-degree on the graph, namely,
\begin{equation}
N_{ij}^{(k)} \leq \Delta^{k-1}. \label{eq:Nijk-limit}
\end{equation}

Hence,
\begin{align}
C(\gamma) &\le \gamma \Delta^{D_{ij}}  \sum_{k=0}^{\infty} (\gamma \Delta) ^k \\
&= \frac{\gamma \Delta^{D_{ij}}}{1-\gamma\Delta} \xrightarrow{\gamma \rightarrow 0} 0.
\end{align}

Therefore,
\begin{align}
R_{ij}(\gamma) &= \log_{\gamma} Y_{ij}(\gamma) \\
&= D_{ij} + \log_{\gamma}\left( S_{ij} + C \right) \xrightarrow{\gamma \rightarrow 0} D_{ij},
\end{align}
which proves Eqn.~\ref{eq:R-limit}.

\subsection{Extension to weighted graphs} \label{sec:weighted-graphs}

Consider now the more general case of directed graphs with positive integer weights. Let the edge from node $j$ to node $i$ have weight $W_{ij}$ with
\footnote{We follow the usual convention for operations involving "$\infty$": $a+\infty=\infty, \, \forall a \in \mathbb{R}\cup \{\infty\}$ and $\gamma^{\infty}=0$ if $0 < \gamma < 1$. }
$$
W_{ij}  \in \mathbb{N}^+ \cup \{\infty\}. 
$$

The length of a path is defined as the sum of the weights of the edges. The distance $D_{ij}$ from $j$ to $i$ is the shortest length of a path from $j$ to $i$. 

Note in the special case of an unweighted graph considered above, ${W_{ij}}=1$ if there is an edge from $j$ to $i$, and ${W_{ij}}=\infty$ otherwise.

Now, define ${\mathbf{X(\gamma)}} = \gamma ^ {\mathbf{W}}$ element-wise, namely,
$$
X_{ij}(\gamma) = \gamma ^ {W_{ij}}.
$$

Then, consider the function
\begin{align}\label{eq:Y-taylor-series-integer}
{\mathbf{Y}(\gamma)} &= \left({\mathbf{I}}-{\mathbf{X}(\gamma)}\right)^{-1} \\
&= {\mathbf{I}} + {\mathbf{X(\gamma)}} +  {\mathbf{X}^2(\gamma)} + \dots .
\end{align}

The Taylor series converges if $\gamma$ is small enough so $\rho(\mathbf{X}(\gamma))<1$.
Note in the special case of an unweighted graph, ${\mathbf{X}(\gamma)} = \gamma {\mathbf{A}}$ is directly proportional to the adjacency matrix. 

As in the unweighted case, the powers of ${\mathbf{X(\gamma)}}$ represent the outcome of multi-step walks, but with an added dimension:
\begin{equation}\label{eq:Mij-gamma-sum}
    {\left[ {{{\mathbf{X}^k(\gamma)}}} \right]_{ij}} = \sum_{d=0}^{\infty} M_{ij}^{(k,d)} \gamma^d ,
\end{equation}
where
$$
M_{ij}^{(k,d)} = {\text{number of distinct walks to get from }}{j}{\text{ to }}{i}{\text{ in }}k{\text{ steps with total length }}d .
$$

Note that the right-hand side of Eqn.~\ref{eq:Mij-gamma-sum} is indeed a finite sum, because, for a given $k$, the length of any $k$-step walk cannot exceed $k$ times the largest integer weight across all edges.

Thus,
\begin{align}
    Y_{ij}(\gamma) &=  \sum_{k=0}^{\infty} \left( \sum_{d=0}^{\infty} M_{ij}^{(k,d)} \gamma^d \right) \label{eq:Yij-series-integer}\\ 
    &= \sum_{d=0}^{\infty} \left( \sum_{k=0}^{\infty}M_{ij}^{(k,d)}\gamma^d \right)  \label{eq:Fubini}\\
    &= \sum_{d=0}^{\infty} N_{ij}^{(d)} \gamma^d \label{eq:Yij-equal-Nij-gamma-integer},
\end{align}
where
$$
N_{ij}^{(d)} = \sum_{k=0}^{\infty}M_{ij}^{(k,d)} = {\text{number of distinct paths to get from }}{j}{\text{ to }}{i}{\text{ with length }}d.
$$

We are allowed to interchange the order of summation because of Fubini's theorem for infinite series: the double series is absolutely convergent because each term of Eqn.~\ref{eq:Yij-series-integer} is positive.

Note the analogy to ($\ref{eq:Y-series}$) above. 
From here the argument proceeds as above. The first non-zero term in the power series occurs for $d=D_{ij}$. The remaining sum is bounded because $N_{ij}^{(d)}$ is bounded from above by $\Delta^{d-1}$. So one finds that for positive-integer-weighted graphs, the R-distance, in its generalized form, again delivers graph distances in the asymptotic limit of small $\gamma$:
\begin{equation}
R_{ij}(\gamma) = \log_{\gamma} {\left[ {\left({\mathbf{I}}-\gamma ^ {\mathbf{W}}\right)^{-1}} \right]}_{ij}   \xrightarrow{\gamma \rightarrow 0} D_{ij}. \label{eq:distance-function-integer}
\end{equation}

\subsection{Positive real weights}

What if the weights are non-integer? For any practical application, one can always discretize the weights with some fine enough resolution $\Delta W$:
\begin{equation}
    W_{ij} \approx V_{ij} \Delta W ,
\end{equation}
where the $V_{ij}$ are integers, and $\Delta W$ is some weight increment smaller than the resolution needed for the problem at hand. Then one can proceed with finding the shortest paths on the resulting integer-weighted graph. Therefore, even in the case of non-integer weights, the R-distance of Eqn~\ref{eq:distance-function-integer} delivers correct graph distances in the limit of small $\gamma$. A more formal proof appears in the Appendix \ref{appendix-real-wts}.

\begin{figure}
  \centering
    \includegraphics[width=1.0\linewidth]{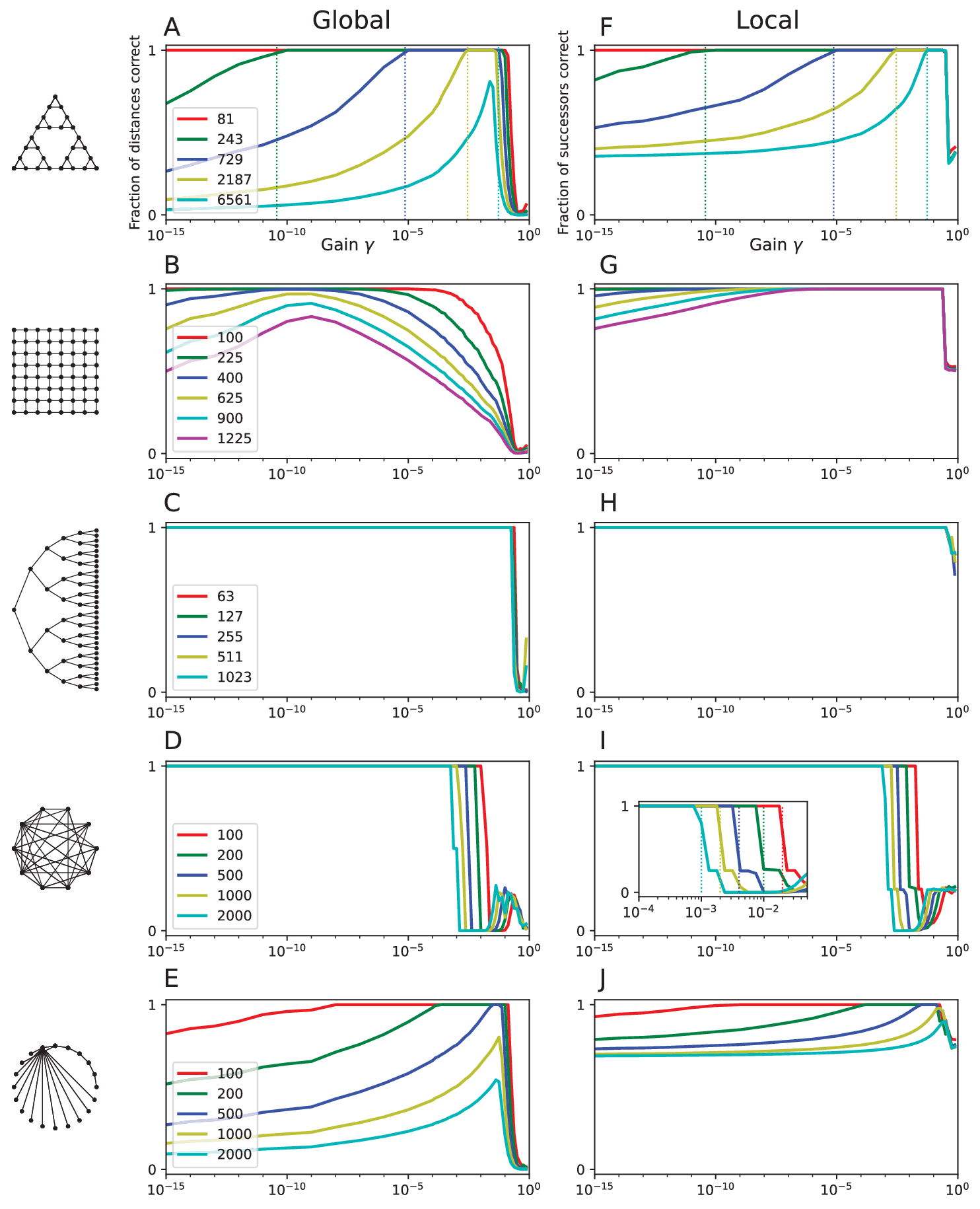}
    \caption{{\bf{Performance of the R-distance.}} Each row represents a family of graphs illustrated by the icon on the left, with node numbers given in the legend. {\bf{Global:}} Global performance: fraction of distances between nodes predicted correctly by Eqn \ref{eq:rounded-distance-function}, as a function of the gain parameter $\gamma$. {\bf{Local:}} Local performance: fraction of successors predicted correctly according to Eqn \ref{eq:descent}. {\bf{A:}} Global performance on graphs representing the Towers-of-Hanoi game. Dotted lines show the cutoff imposed by machine precision, according to Eqn \ref{eq:gamma-precision}. {\bf{F:}} as in A, but for local performance. Note the valid range of $\gamma$ extends further to the right. {\bf{B, G}} As in A and E, but for grid graphs: The nodes lie on a square Cartesian grid and each node connects to the 4 nearest neighbors. {\bf{C, H}} Same, but for binary tree graphs, in which each node connects to 2 child nodes up to a certain number of levels. {\bf{D, I}} Same, but for dense random graphs, in which each pair of nodes is connected with probability 0.5. {\bf{I inset}}: Dotted lines indicate the critical gain, Eqn~\ref{eq:critical-gain}. {\bf{E, J}} Same, but for graphs following a power law degree distribution (exponent = 3).}
\label{fig:performance}
\end{figure}

\section{The R-distance at finite gain} \label{sec:finite}

For any practical application, one needs to compute the R-distance using some non-zero gain $\gamma$. Furthermore, as is apparent from the power series (\ref{eq:Y-series-trunc}), the calculation of a long path involves high powers of $\gamma$. Thus one wants to keep $\gamma$ as large as possible, so these terms don't get swamped by numerical imprecision. Here we consider the constraints on $\gamma$ to deliver a useful distance function. We focus on the case of unweighted graphs (Eqns~\ref{eq:adjacency} and~\ref{eq:R-defined}).

\subsection{Global distance function} \label{sec:global-distance}

We say that the R-distance is \textbf{globally correct} if, when rounded up to the nearest integer, it delivers the correct distances between any two nodes on the graph:
\begin{equation}
    \left\lceil R_{ij}(\gamma) \right\rceil = \left\lceil \log_{\gamma} {\left[ {\left({\mathbf{I}}-\gamma {\mathbf{A}}\right)^{-1}} \right]}_{ij} \right\rceil = D_{ij} . 
\label{eq:rounded-distance-function}
\end{equation}

An equivalent condition is 
\begin{equation}
    \gamma^{D_{ij}} \le Y_{ij}(\gamma) < \gamma^{D_{ij}-1},\, \forall \, i,j,
    \label{eq:Y-limits}
\end{equation}
where
\begin{equation}
    {\bf{Y}(\gamma)} = \left({\bf{I}} - \gamma {\bf{A}} \right) ^ {-1}.
\end{equation}

Starting with the arguments in Section \ref{sec:asymptotic}, one is led to several interesting bounds on the value of $\gamma$. We illustrate these conditions with the numerical experiments of Figure~\ref{fig:performance}A-E, where the fraction of correct graph distances across all node pairs $i,j$ is plotted against the value of the gain $\gamma$.

\subsubsection{Critical gain}

The Taylor expansion in Eqn \ref{eq:inverse-series} has a convergence radius of 1. So if the spectral radius of $\gamma {\mathbf{A}}$ exceeds 1, this expansion no longer holds, and thus ${\mathbf{Y}}(\gamma)$ no longer represents the lengths of paths on the graph (Eqn \ref{eq:Y-series}). Therefore an \emph{upper bound} on $\gamma$ is given by the critical gain $\gamma_{\rm{c}}$: 
\begin{equation}
\gamma < \gamma_{\rm{c}} \equiv \frac{1}{\rho({\bf{A}})} ,\label{eq:critical-gain}
\end{equation}
where $\rho({\bf{A}})$ is the spectral radius of ${\bf{A}}$, namely, the maximum of the absolute values of the eigenvalues of  $\mathbf{A}$.

For unweighted graphs, the spectral radius is closely related to the number of edges per node~\citep{brouwerSpectraGraphs2012}. In Figure \ref{fig:performance}, one sees that the performance of the R-distance drops off sharply at large values of $\gamma$ for every graph type: This is the critical gain for that graph. For the random dense graphs (Fig \ref{fig:performance}D), the number of edges per node grows with the number of nodes, so the upper cutoff value of $\gamma$ decreases with the size of the graph.

However, condition (\ref{eq:critical-gain}) does not guarantee that the R-distance is globally correct, as seen in the following example:
    
\begin{example}\label{example:path-graph}
    Consider the undirected path graph with three vertices, whose adjacency matrix $\mathbf{A}$ is given by:
    \begin{equation}
        \mathbf{A} = \begin{bmatrix}
            0 & 1 & 0\\
            1 & 0 & 1\\
            0 & 1 & 0
        \end{bmatrix}.
    \end{equation}

    Then,
    \begin{equation}
        (\mathbf{I}-\gamma \mathbf{A})^{-1} = \frac{1}{1-2\gamma^2}\begin{bmatrix}
            1-\gamma^2 & \gamma & \gamma^2\\
            \gamma & 1 & \gamma\\
            \gamma^2 & \gamma & 1-\gamma^2
        \end{bmatrix}.
    \end{equation}
    
    The eigenvalues of $\mathbf{A}$ are $\lambda=0,\pm\sqrt{2}$, so the critical gain is $\gamma_{\rm{c}}=1/\sqrt{2}$. Let us consider the R-distance between vertex $1$ and $2$, namely, $\log_{\gamma}[(\mathbf{I}-\gamma \mathbf{A})^{-1}]_{12}$. As $\gamma$ approaches the critical gain from below, we have    
    \begin{equation}
        \lim_{\gamma \rightarrow \gamma_{\rm{c}}^-}\log_{\gamma}[(\mathbf{I}-\gamma \mathbf{A})^{-1}]_{12} = \lim_{\gamma \rightarrow \gamma_{\rm{c}}^-}\log_{\gamma}\frac{\gamma}{1-2\gamma^2} = -\infty,
    \end{equation}
    whereas the correct answer is $D_{12}=1$. Therefore, there exists a finite range of $\gamma$ below $\gamma_{\rm{c}}$ for which the R-distance is not globally correct.
    \qed
\end{example}

\subsubsection{An upper bound on the gain from redundant paths}

The critical gain (\ref{eq:critical-gain}) places an upper bound on $\gamma$, but for certain graphs another upper bound is much more restrictive. 

\begin{theorem}\label{thm:global-necessary-conditions}
    For any unweighted directed graph, the R-distance is globally correct (Eqn~\ref{eq:rounded-distance-function}) only if the following inequality is satisfied:
    \begin{equation}
        \gamma < \frac{1}{\max_{i,j} S_{ij}},
        \label{eq:redundant-paths-bound}
    \end{equation}
    where $S_{ij}$ is the number of distinct shortest paths from $j$ to $i$.
\end{theorem}

\begin{proof}
    For a globally correct R-distance, $\gamma$ must satisfy Eqn \ref{eq:Y-limits}. 
    Because $Y_{ij}(\gamma) \ge S_{ij} \gamma^{D_{ij}}$ (Eqn~\ref{eq:Y-series-trunc}), this requires that
    \begin{equation}
        S_{ij} \gamma^{D_{ij}} < \gamma^{D_{ij}-1},\, \forall \, i,j.
    \end{equation}
    
    That is,
    \begin{equation}
        \gamma < \frac{1}{S_{ij}} ,\, \forall \, i,j.
    \end{equation} 
\end{proof}

Theorem~\ref{thm:exponential-bound} implies that for certain graph types, the upper bound on $\gamma$ declines exponentially with the size of the graph. 

\begin{example}
    Consider a square grid graph in which the nodes lie on a square Cartesian grid and each node connects to the 4 nearest neighbors (Fig~\ref{fig:performance}B). If the grid measures $M$ nodes on a side, then the graph diameter is $d = 2M-2$, and the largest degree is $\Delta = 4$. Taking nodes $i,j$ to be opposite corners of the grid, the distance is $D_{ij}=2M-2$, but there are a huge number of redundant shortest paths with that same distance: 
    \begin{equation}
        N_{ij}^{(D_{ij})} = \binom{2M-2}{M-1} > 2^{M-1} = 2^{d/2}.
    \end{equation}
    That number grows exponentially with the diameter of the graph.
    \qed
\end{example}

The effect can be seen in the numerical results of Figure~\ref{fig:performance}B, where the accuracy of the R-distance declines steeply at large $\gamma$ and more so the larger the graph. By contrast, this high degree of redundancy among shortest paths does not appear in the other graph types tested here (Fig \ref{fig:performance}A,C,D,E). The binary tree graphs allow the greatest range of $\gamma$ for a globally correct R-distance, with the only practical constraint being the critical gain (Fig \ref{fig:performance}C). These trees don't suffer from redundant paths, and the diameter of the graph grows only logarithmically with the number of nodes.

\subsubsection{A sufficient condition for the gain}

Next, we provide a \emph{sufficient condition} for a globally correct R-distance.

\begin{theorem} \label{thm:exponential-bound}
    For any unweighted directed graph,
    the R-distance is globally correct (Eqn~\ref{eq:rounded-distance-function}), provided that
    \begin{equation}
        0 < \gamma < \frac{1}{\Delta + \Delta^{d-1}}, 
    \end{equation}
    where $d=\max_{i,j}D_{ij}$ is the diameter of the graph and $\Delta=\max_j(\sum_iA_{ij})$ is the largest out-degree among all vertices.
\end{theorem}
\begin{proof} 
    Starting with Eqns~\ref{eq:Y-series-trunc} and \ref{eq:Nijk-limit} from Section~\ref{subsec:unweighted-directed-graph}, we obtain two inequalities: 
    \begin{align}
        \gamma^{D_{ij}} \le Y_{ij}(\gamma) \le \sum_{k=D_{ij}}^{\infty}\gamma^k\Delta^{k-1} = \frac{\gamma^{D_{ij}}\Delta^{D_{ij}-1}}{1-\gamma \Delta}.
        \label{eq:Y-bracket}
    \end{align}
    
    The left inequality arises because  $S_{ij} \ge 1$ in Eqn~\ref{eq:Y-series-trunc}. The right inequality comes from inserting Eqn.~\ref{eq:Nijk-limit} in the right-hand side of Eqn.~\ref{eq:Y-series-trunc}.

    Now choose $\gamma$ small enough such that for any $i,j$,
    \begin{equation}
        \frac{\gamma^{D_{ij}}\Delta^{D_{ij}-1}}{1-\gamma \Delta} < \gamma^{D_{ij}-1}.
    \end{equation}
    
    This is satisfied if 
    \begin{equation}
        \gamma < \frac{1}{\Delta + \Delta^{d-1}}.
    \end{equation}
    
    Then Eqn.~\ref{eq:Y-bracket} leads directly to Eqn.~\ref{eq:Y-limits}.
\end{proof}

\subsubsection{Bounds for the gain on tree graphs}

Next we consider a special case: the tree, which is a very common class of graphs. On a tree, there is only one shortest path between any two nodes, so Theorem \ref{thm:global-necessary-conditions} does not constrain the gain. 
Thus, one might wonder if there exists a more generous range over which the 
R-distance is globally correct. Indeed, in the case of the tree, one can obtain a sufficient upper bound on $\gamma$ of order $\mathcal{O}(1/\sqrt{\Delta d})$, which declines only as a power function of the diameter. 

\begin{theorem} \label{thm:power-bound-induction}
    For any undirected tree graph 
    with maximum vertex degree $\Delta$ and diameter $d$, the R-distance is globally correct (Eqn~\ref{eq:rounded-distance-function}), provided that
    \begin{equation}
        0 < \gamma < \frac{-1+\sqrt{1+4\Delta(d+2)}}{2\Delta(d+2)}.
    \end{equation}
\end{theorem}

\begin{proof}
    See Appendix \ref{appendix-tree-bound}.
\end{proof}

The above upper bound depends on the diameter, which is not readily available if one is given only the adjacency matrix. Nonetheless, on can bound $d$ by the number of vertices $N$, which results in an upper bound of order $\mathcal{O}(1/\sqrt{N\Delta})$, independent of the diameter. 

We note that additional bounds on $\gamma$ may be obtained by algebraic methods, exploiting the connection between the matrix $\bf{Y}$ (Eqn~\ref{eq:Y-defined}) and the characteristic polynomial of the graph.

\subsection{Local distance function}

Perhaps the most common use of an all-pairs distance matrix $D_{ij}$ is to find the actual shortest paths between any two nodes on the graph. A simple greedy-descent algorithm accomplishes this. Say the goal node is $g$. Starting from node $i$, find all the nodes connected to it. Within that set of successor nodes ${\mathcal{S}}(i)$, choose the node $j$ that has the shortest distance $D_{gj}$ to the goal:
\begin{equation}
i \gets \underset{j \in {\mathcal{S}}(i)}{\argmin} \, D_{gj}. \label{eq:descent}
\end{equation}

Then iterate this step until the goal is reached.

If one uses the R-distance function $R_{gj}$ for this algorithm instead of the exact graph distances $D_{gj}$, one will of course find the shortest paths if the R-distance is globally accurate across the entire graph (Eqn \ref{eq:rounded-distance-function}, Fig \ref{fig:performance}A-E). However, the R-distance turns out to yield shortest paths even when it is not globally accurate. Over a wide range of $\gamma$, the R-distance, when applied in Eqn~\ref{eq:descent}, chooses the correct successor node for all pairs of start and goal nodes (Fig \ref{fig:performance}F-J). In those cases one can say the R-distance is \emph{locally correct}.

The effect is most pronounced for the example of grid graphs (Fig~\ref{fig:performance}G). Here, a wide range of $\gamma$ supports locally correct performance, even when there exists no $\gamma$ that produces globally correct distances (Fig~\ref{fig:performance}B). 
The constraint on $\gamma$ from redundant paths (Eqn~\ref{eq:redundant-paths-bound}) largely drops away, such that values all the way up to the critical gain $\gamma_{\rm{c}}$ support greedy-descent on the R-distance.
One can understand this by considering Eqn~\ref{eq:Y-series-trunc}: In the power series for $Y_{gj}$, the first non-vanishing term is $S_{gj} \gamma^{D_{gj}}$. 
Comparing nodes $j$ that are all neighbors of $i$, the number of redundant shortest paths to the goal $S_{gj}$ will be quite similar. 
That leaves the distance-dependent term $\gamma^{D_{gj}}$ to dominate the R-distance. This is why making local comparisons of the R-distance among nearby points still identifies the node nearest the goal correctly.

For most of the graphs tested in Figure~\ref{fig:performance} the R-distance is locally correct as long as $\gamma$ is chosen just below the critical gain (Fig~\ref{fig:performance}I inset). 

\begin{figure}
  \centering
    \includegraphics[width=1.0\linewidth]{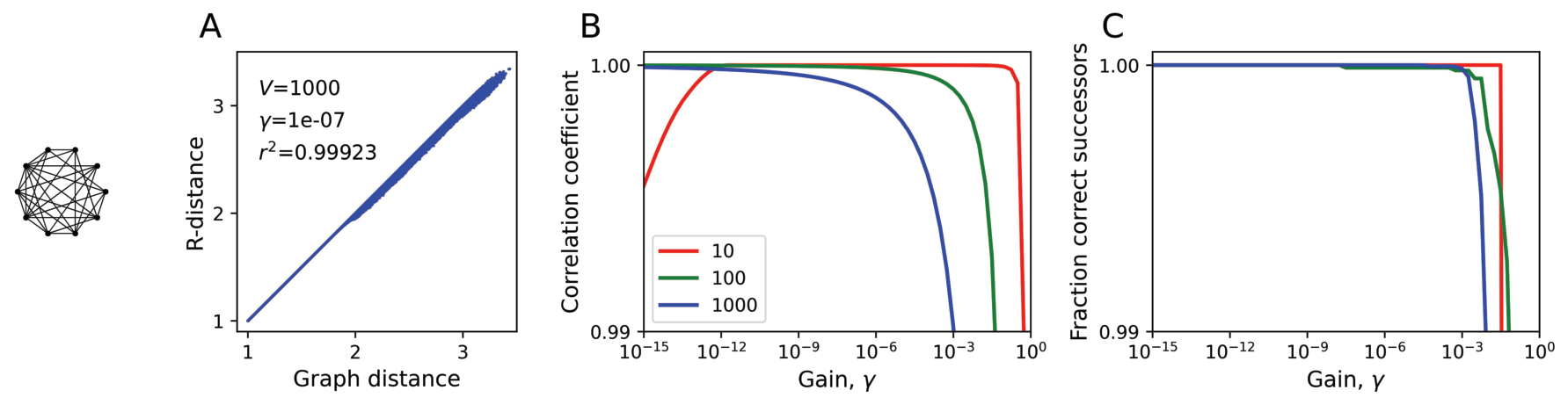}
    \caption{{\bf{The R-distance on weighted graphs.}} Results on a dense graph (as in Fig~\ref{fig:performance}D,I) with random positive edge weights. {\bf{A:}} The R-distance plotted against the true graph distance. $V=1000$ vertices, $p=0.5$ edge probability, edge weights range from 1 to 100 with log-uniform distribution. $\gamma = 10^{-7}$. Note high correlation coefficient $r^2$. {\bf{B:}} Global performance. The correlation coefficient $r^2$ between R-distance and true graph distance, for graphs of various sizes ($V$ in legend) as a function of the gain $\gamma$. Note y-axis scale is close to 1. {\bf{C:}} Local performance. The fraction of correct successors predicted by the R-distance. At $\gamma < 10^{-8}$ all 1 million successors in the $V=1000$ graph are predicted correctly.}
\label{fig:positive-weights}
\end{figure}

\subsubsection{A sufficient condition for greedy descent}
Consider an agent pursuing a goal $g$ by performing greedy descent on the R-distance according to Eqn~\ref{eq:descent}.
If the R-distance has a local minimum, the agent will get stuck, cycling endlessly between neighboring nodes.
The following condition guarantees the absence of such minima, so the agent reaches the goal, although not necessarily by the shortest route.

\begin{theorem}    
\label{thm:largest-degree-bound}
    For any 
    graph $G$ with maximum vertex out-degree $\Delta$, greedy descent on the R-distance (Eqn~\ref{eq:descent}) will lead to the goal, if $0<\gamma<1/\Delta$
\end{theorem}

\begin{proof}
    Greedy descent on the R-distance $R_{gi}$ to the goal $g$ is equivalent to greedy ascent on $Y_{gj}$ (Eqn~\ref{eq:R-defined}). 
    We want to show that every node $i \ne g$ has a neighboring node $j$ where $Y_{gj}>Y_{gi}$. Then the agent will move to $j$ and iterate until it arrives at $g$.
    
    Here, $Y_{gi}$ is the $g$-th row of the matrix $(\mathbf{I} - \gamma \mathbf{A})^{-1}$. By evaluating $\mathbf{Y} (\mathbf{I} - \gamma \mathbf{A}) = \mathbf{I}$ one finds for $i \ne g$:
    \begin{equation}
        Y_{gi} = \gamma \sum_j Y_{gj} A_{ji} < \frac{1}{\Delta_i} \sum_{j:j \sim i} Y_{gj} = \left< Y_{gj} \right>_{j:j \sim i},
    \end{equation}
    where $\Delta_i$ is the out-degree of node $i$, and the inequality follows from the condition that $\gamma<1/\Delta_i$. In words, $Y_{gi}$ is smaller than the average $Y_{gj}$ at all the neighbors of $i$. Because all the $Y_{gj}$ are non-negative, at least one of them must be larger than $Y_{gi}$. Furthermore, the unique maximum can only be obtained at the goal node $i=g$.    
\end{proof}

\begin{corollary}
    Under the assumption of theorem~\ref{thm:largest-degree-bound}, the path generated by the greedy algorithm does not contain any cycles. In particular, when the graph is a tree, this is also the shortest path.
\end{corollary}
\begin{proof}
    Since the inequalities in the proof of Theorem~\ref{thm:largest-degree-bound} are all strict, we can not have a cycle in the greedy path. For the second claim, we note that on a tree, the only path without cycles is the unique shortest path.
\end{proof}

\subsection{R-distance on weighted graphs}
So far, we have considered unweighted graphs where the edges are either present or not. However, as derived in Section~\ref{sec:weighted-graphs}, the R-distance should be a useful measure also on weighted graphs, as long as the edge weights are positive real numbers. On a weighted graph, the length of a path is the sum of the weights of its edges, and the shortest path is the one with the smallest total weight. 

Figure~\ref{fig:positive-weights} illustrates the relation of the R-distance to the true distance on a dense graph with random positive weights, whose values span two orders of magnitude. Because the distance can take on any real value, one cannot expect the R-distance to be exact. However, it can be tightly correlated to the real distance (Fig~\ref{fig:positive-weights}A), and over a wide range of gain values $\gamma$ the correlation is almost perfect (Fig~\ref{fig:positive-weights}B). Furthermore, over a wide range of $\gamma$, the R-distance is locally correct, in that it identifies every correct successor node for the shortest path (Fig~\ref{fig:positive-weights}C).

\begin{figure}
  \centering
    \includegraphics[width=1.0\linewidth]{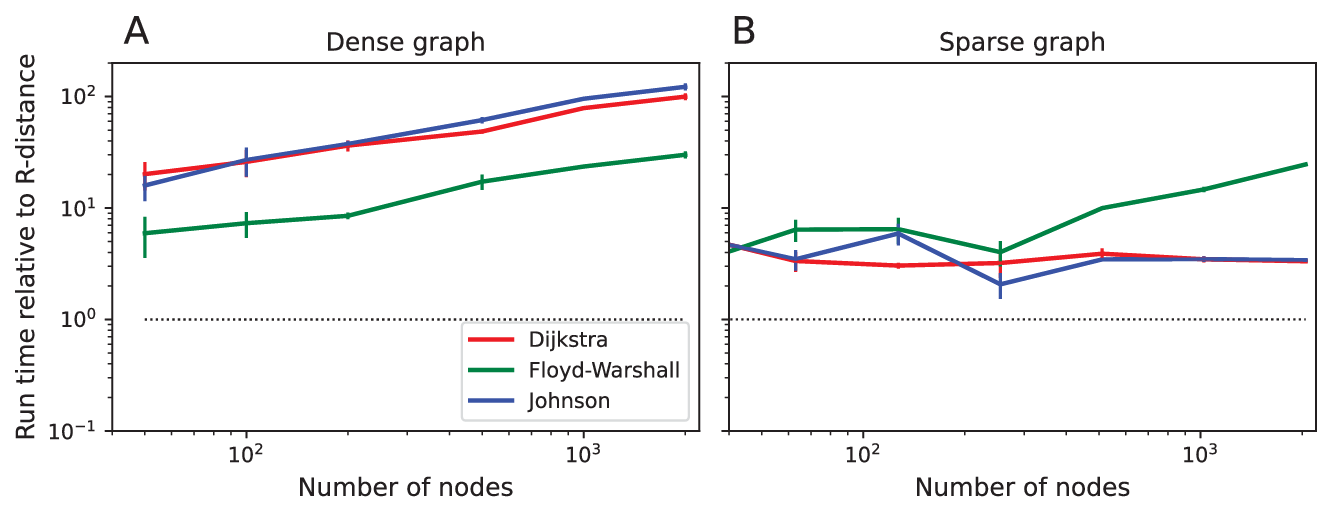}    \caption{{\bf{Run times to compute distance functions.}} The run times for 3 conventional APSP algorithms, compared to that for the R-distance. The algorithms are Floyd-Warshall, Dijkstra, and Johnson, as implemented in the Python package \texttt{scipy}. {\bf{A:}} Dense graphs (random unweighted with edge probability 0.5) with varying node numbers. All run times are divided by that for the R-distance. Dotted line: equality. Mean over 25 runs, bars indicate standard error. {\bf{B:}} As in (A) for sparse graphs (binary tree). Note for all these graphs the conventional APSP algorithms are slower than the R-distance.}
\label{fig:run-times}
\end{figure}

\section{Digital computation} \label{sec:digital}

While we have introduced this solution to the APSP problem as uniquely suited to analog networks (Fig~\ref{fig:circuit}), there may be situations even on digital computers in which computing the R-distance is preferable to the conventional APSP algorithms. Here we encounter another constraint on the value of the gain $\gamma$: the machine precision.

\subsection{Machine precision} \label{sec:machine-precision}

Suppose the largest distance on the graph is $d$ and there is just one path with that distance. Then for that pair of nodes $Y_{ij}(\gamma)=\gamma^{d}$ (Eqn \ref{eq:Y-series}). This must be a valid number, so it should exceed the smallest representable number $\delta$. This results in a \emph{lower bound} for $\gamma$, namely,
\begin{equation}
\gamma > \delta ^ {1/d} \label{eq:gamma-precision}.
\end{equation}

In the performance measures of Figure~\ref{fig:performance} the effect of machine precision can be seen clearly for the Towers-of-Hanoi graphs, where the low-$\gamma$ cutoff is determined precisely by the longest distance on the graph (Fig \ref{fig:performance}A). 

If a graph contains some long distances, then this lower bound may conflict with the upper bound given by the critical gain. To satisfy both, one requires
\begin{equation}
d < \frac{\log \delta}{\log \gamma_{\rm{c}}} .  \label{eq:longest-distance}
\end{equation}

For example, in double precision arithmetic (IEEE 754 standard) $\delta \approx 5 \times 10^{-324}$. Suppose the graph has a spectral radius of $\lambda_{\rm{max}} \approx 10$, then $\gamma_{\rm{c}} \approx 0.1$, which sets a limit on the maximal distance $d<323$. If the graph includes larger distances, then the R-distance function cannot be globally correct for any value of $\gamma$. One can see this conflict for the larger Tower-of-Hanoi graphs (Fig \ref{fig:performance}A) and in certain Power law graphs (Fig \ref{fig:performance}E). 

\subsection{Complexity}

The time complexity of the R-distance (Eqn \ref{eq:R-defined}) is dominated by the matrix inversion, which for a graph with $N$ vertices has complexity $\mathcal{O}(N^\omega)$ (currently $\omega=2.373$). This improves on published algorithms for APSP on dense directed graphs, which have a higher polynomial dependence on $N$~\cite{zwickExactAndApproximate2001}. For example, Takaoka's algorithm for graphs with integer weights \cite{takaokaSubcubicCostAlgorithms1998} runs in $\mathcal{O}(M^{1/3}N^{(6+\omega)/3})$ where $M$ is the largest weight. Thus the R-distance proposed here has a time-complexity better than all known algorithms for dense directed graphs~\citep{chanMoreAlgorithmsAllPairs2010}. 

However, this theoretical advantage comes with some caveats. For one, the R-distance can be globally correct only if the critical gain exceeds the lower bound imposed by the machine precision and the longest distance (Eqn \ref{eq:longest-distance}). 
As the number of vertices $N$ grows, the largest distance on the graph will grow as well, necessitating greater bit depth for the computation. As $N$ goes to infinity, one would need infinite machine precision. Effectively the computation of R-distance pushes part of the complexity from time into space. It shares that characteristic with some other APSP algorithms that presume infinite machine precision~\citep{yuvalAlgorithmFindingAll1976}. 

Second, from a practical perspective, there is no current implementation of matrix inversion or even matrix multiplication that runs in $\mathcal{O}(N^\omega)$. Some have expressed doubts whether a machine will ever be built on which the Coppersmith-Winograd algorithm \cite{coppersmithMatrixMultiplicationArithmetic1990} offers a benefit over the schoolbook $\mathcal{O}(N^3)$ multiplication \cite{robinsonOptimalAlgorithmMatrix2005}. Certainly the popular scientific programming platforms all invert dense matrices with an $\mathcal{O}(N^3)$ algorithm.

Nonetheless, we have found in practice that the R-distance is locally correct on many graphs up to thousands of nodes in size (Figs \ref{fig:performance}, \ref{fig:positive-weights}). Furthermore, in at least one popular scientific programming environment (Python \texttt{scipy}) its evaluation is more than $10 \times$ faster than the Floyd-Warshall algorithm on dense graphs (Fig \ref{fig:run-times}A). Even on sparse graphs, the R-distance is considerably faster than Dijkstra's and Johnson's algorithms, which were designed for sparse graphs (Fig \ref{fig:run-times}B). This can be attributed to the extensive effort that has gone into optimizing the routines for matrix inversion.

\section{Discussion}

\subsection{Related work}

The central observation here is that one can obtain all the pairwise shortest distances on a graph directly from the resolvent function $\left( {\mathbf{I}} - \gamma {\mathbf{A}} \right) ^ {-1}$ of its adjacency matrix (Section \ref{sec:asymptotic}). The approach is valid over a wide range of graph types and sizes (Section \ref{sec:finite}) and performs efficiently compared to conventional APSP algorithms (Section \ref{sec:digital}). Unlike other known algorithms, computation of the R-distance maps naturally onto an analog neural network (Section~\ref{sec:network}). 

The resolvent function has played a important role in the analysis of graphs for some time~\citep{katzNewStatusIndex1953,benziMatrixFunctionsNetwork2020}, and the developments here may allow another interpretation of those measures. For example, the ``resolvent-based total communicability'' has been defined~\citep{benziTotalCommunicabilityCentrality2013} as
\begin{equation}
    C_r({\mathbf{A}})=\sum_{i=1}^n \sum_{j=1}^n \left[\left( {\mathbf{I}} - \gamma {\mathbf{A}} \right) ^ {-1} \right] _{ij}.
\end{equation}    

The authors recommend setting $\gamma$ just below the critical value, $\gamma = 0.85 \, \gamma_{\rm{c}}$, without much justification. From Figure~\ref{fig:performance} we see that this value is too close to the critical value to produce a globally correct distance function on most graphs. Hence the meaning of this measure will depend strongly on the graph type. If we choose $\gamma$ somewhat smaller, say $\gamma = 0.01 \gamma_{\rm{c}}$, then we can use Eqns~\ref{eq:R-defined} and \ref{eq:R-limit} to reinterpret the communicability. In that case, $C_r$ becomes a simple function of all the graph distances $D_{ij}$:
\begin{equation}
    C_r=\sum_{i=1}^n \sum_{j=1}^n \gamma ^ {D_{ij}}.
\end{equation}    

This gives a more concrete understanding of the concept of ``communicability''.  

With the goal of determining all shortest paths, a recent report~\citep{steinerbergerSpectralApproachShortest2021} attempts a solution based on the Laplacian matrix, which is related to the resolvent, but lacks the degree of freedom given by $\gamma$. That method is computationally expensive, gives provably correct results only on trees, and fails even on elementary graphs~\citep{jourdanSpectralMethodFinding2021}. 
Another study used the resolvent function to estimate graph distances~\citep{baramIntuitivePlanningGlobal2018}, but again with a fixed value of $\gamma = 0.85 \, \gamma_{\rm{c}}$, and found empirically that the estimates were approximately correct on some graphs. Again, the developments presented here explain why and how one should choose $\gamma$ to obtain a useful distance function on diverse graph types.

\subsection{Applications}

On digital computers, the R-distance probably offers limited benefits over conventional APSP algorithms. While it does execute faster than the plain-vanilla implementations of APSP on graphs with several thousand nodes (Fig~\ref{fig:run-times}), any truly time-critical application will likely involve huge graphs. If the longest distance on the graph exceeds a few hundred nodes, the R-distance algorithm will run into the machine precision limit (Eqn~\ref{eq:longest-distance}). Furthermore, if one optimizes the Floyd-Warshall algorithm for large node numbers and parallel GPU architectures, one can achieve efficiencies similar to the high-performance routines for matrix inversion~\citep{saoScalableAllpairsShortest2021}.

By contrast, the R-distance function seems perfectly adapted for \emph{analog computation}. In the circuit of Figure~\ref{fig:circuit}, the knowledge of the graph is embodied in the feedback connection strengths, and the network output can be used to navigate on the graph. For example, one can envision such a circuit in a robot control system. Here the graph represents a map of the environment with navigable paths between nodes. The robot specifies the goal location, by measuring unit $g$'s output, and its current location, by setting unit $i$'s input to 1. Then it queries the analog network for the next step to take towards the goal (Eqn~\ref{eq:descent}). 
Often a robot must navigate in an uncertain environment, and update its map when a new obstacle is encountered~\citep{koenigFastReplanningNavigation2005}. Then the shortest paths must be recomputed dynamically. For this purpose, the circuit of Figure~\ref{fig:circuit} is ``always on'' and responds instantaneously to any changes in the connection strengths $A_{ij}$ or in the current input node. If implemented in analog electronics, the settling time for the new result could be microseconds or less. The low component count, with just a few transistors per node, may allow miniaturization of such a circuit down to millimeter-sized robots. The power needs would be minuscule compared to running a conventional APSP algorithm on a digital processor.

Another suitable analog network is the brain. Animal brains must routinely solve problems that reduce to search on a graph. Take the explicit instance of spatial navigation: An animal explores a new physical environment, learns the available paths between points, then navigates towards chosen goal locations on that graph~\citep{tolman1948cognitive}. These functions are essential to insects, with brains of $\sim 10^{5}$ neurons, as much as to humans with $\sim 10^{10}$~\citep{webbNeuralMechanismsInsect2016,epsteinCognitiveMapHumans2017}. One suspects that evolution has found an efficient way to handle graph search using a neural network. Indeed, the circuit in Figure~\ref{fig:circuit} is a common motif found in animal brains~\citep{dayan_theoretical_2001,muller_hippocampus_1996,douglasRecurrentNeuronalCircuits2007}. Here each active unit is a neuron or a group of neurons, and the connections are recurrent synapses among those neurons. Thus the R-distance computed by such a neural circuit could serve as the goal signal that guides the animal's navigation. 

A key concern in analog computation is noise. Neurons as well as transistors operate with some finite signal-to-noise ratio. This noise will set a limit on the accuracy of the shortest-path search, for example when the greedy descent agent must discriminate which of two signals is larger (Eqn \ref{eq:descent}). The effect is much like the role of machine precision on digital systems (Sec~\ref{sec:machine-precision}). 
For example, human subjects can discriminate between two stimuli when they differ in intensity by 0.1-1\%. So let us say conservatively that the entire APSP network has a relative precision of $\delta=0.01$.
Then by the arguments of section \ref{sec:machine-precision}, the R-distance is globally correct only if $\gamma > \delta^{1/d}$, where $d$ is the largest distance on the graph. But the gain must also be less than the critical value (Eqn~\ref{eq:critical-gain}). On a typical graph with node degree 3, that would mean 
$$
d < \frac{1}{\log_{\delta} \gamma_{\rm{c}}} \approx 4 ,
$$
which limits the application to rather small graphs. On the other hand, it has been shown that the \emph{local} function of the R-distance, as used during greedy descent (Eqn~\ref{eq:descent}), works over a much wider range for distances up to 12 steps or more, depending on the graph type~\citep{zhangEndotaxisNeuromorphicAlgorithm2024}.

Of course a biological system must solve several additional problems: how to learn the synaptic connections $A_{ij}$ during spatial exploration; how to learn about interesting locations on the graph and store them for future navigation; how to maintain multiple maps of different environments. This is an area of intense investigation~\citep{sosaNavigatingReward2021,peerStructuringKnowledgeCognitive2021}. A recent report proposes and end-to-end solution that implements all these functions on the basis of the R-distance~\citep{zhangEndotaxisNeuromorphicAlgorithm2024}.

\newpage
\appendix
\section{Appendix} \label{appendix}

\subsection{Asymptotic limit of the R-distance with non-integer weights}  \label{appendix-real-wts}
Here we show that Eqn~\ref{eq:distance-function-integer} holds even when the weights $W_{ij}$ are positive real numbers, not necessarily integers, namely,
\begin{equation*}
    W_{ij} \in \mathbb{R}^+ \cup \{\infty\}.
\end{equation*}

The arguments below largely follow those for positive integer weights with an additional complication: the length of a walk is not necessarily an integer, preventing us from directly using the geometric series argument as above. To deal with this, we can enumerate all possible lengths of a walk in ascending order: Let $\mathcal{D}_k$ denote the set of all possible lengths of a $k$-step walk. Since $\mathcal{D}_k$ is a finite set, we know that the set of all possible walk lengths: $\mathcal{D}=\cup_{k=0}^{\infty}\mathcal{D}_k$ is a countable subset of $\mathbb{R}^+\cup\{0,\infty\}$. Since $\mathcal{D}$ is a well-ordered set with the usual ordering "$\le$", we can enumerate $\mathcal{D}$ in ascending order:
\begin{equation*}
    \mathcal{D} = \{d_m\}_{m=0}^{\infty}, \, \text{where $0=d_0<d_1<\cdots$}.
\end{equation*}

Now, as in Eqn.~\ref{eq:Yij-series-integer} to \ref{eq:Yij-equal-Nij-gamma-integer}:
\begin{align}\label{eq:Y-series-positive-weight}
    Y_{ij}(\gamma) &= \sum_{m=0}^{\infty} N_{ij}^{(d_m)} \gamma^{d_m}\\
    &=  \gamma^{D_{ij}}\left( N_{ij}^{(D_{ij})}+\sum_{m:d_m>D_{ij}}N_{ij}^{(d_m)}\gamma^{d_m-D_{ij}} \right)\\
    &= \gamma^{D_{ij}}\left[ N_{ij}^{(D_{ij})}+C(\gamma) \right],
\end{align}
where
$$
N_{ij}^{(d_m)} = \sum_{k=0}^{\infty}M_{ij}^{(k,d_m)} = {\text{number of distinct walks to get from }}{j}{\text{ to }}{i}{\text{ with length }}d_m.
$$

As in the positive integer case, the Taylor series in Eqn.~\ref{eq:Y-taylor-series-integer} converges if $0<\gamma<\delta$ for some sufficiently small $0 <\delta < 1$.

Now, to bound $C(\gamma)$ using a geometric series, we bound the sum of consecutive terms for which the exponent of $\gamma$: $(d_m-D_{ij})$, lies in the integer interval $[l, l+1)$, $l=0,1,2,\cdots$. Let $W_{\min}$ be the minimum weight among all the edges. Then we have
\begin{align}
    \sum_{m:l \le d_m-D_{ij} < l+1} N_{ij}^{(d_m)}\gamma^{d_m-D_{ij}} &\le \sum_{m:d_m-D_{ij} \le l+1} N_{ij}^{(d_m)} \gamma^l\\
    &\le \gamma^l (N+1)^{\tfrac{D_{ij}+l+1}{W_{\min}}},
\end{align}
where the second inequality arises because: any walk from $i$ to $j$ of length at most $(D_{ij}+l+1)$ has at most $\left (\floor \left [\tfrac{D_{ij}+1+l}{W_{\min}} \right ]\right )$ intermediate vertices, excluding $i,j$; and, at each intermediate vertex, there are at most $(N+1)$ options: $N$ different vertices or a dummy ``null'' appended at the tail of short walks.

Thus, we have:
\begin{align}
    0 \le C(\gamma) &\le \sum_{l=0}^{\infty} \gamma^l (N+1)^{\tfrac{D_{ij}+l+1}{W_{\min}}}\\
    &\le \frac{(N+1)^{\tfrac{D_{ij}+1}{W_{\min}}}}{1-\gamma(N+1)^{\tfrac{1}{W_{\min}}}} \xrightarrow{\gamma\rightarrow 0} (N+1)^{\tfrac{D_{ij}+1}{W_{\min}}}.
\end{align}

Hence,
\begin{equation}
    D_{ij}+\frac{\log N_{ij}^{(D_{ij})}}{\log \gamma}\le {\frac{\log Y_{ij}(\gamma)}{\log \gamma}} \le D_{ij} + \frac{\log\left [ N_{ij}^{(D_{ij})} + C(\gamma)\right ]}{\log \gamma}.
\end{equation}

Therefore, by the sandwich theorem, in the limit of small $\gamma$, the R-distance function:
\begin{equation}
R_{ij}(\gamma) = {\frac{\log Y_{ij}(\gamma)}{\log \gamma}} \xrightarrow{\gamma \rightarrow 0}  D_{ij}. \label{eq:distance function-positive}
\end{equation}

\subsection{Proof of Theorem~\ref{thm:power-bound-induction}} \label{appendix-tree-bound}
\begin{proof}
    Note that on trees, there is a unique shortest path connection any pair of vertices. Also, any path connecting a pair of vertices $i,j$ has length $(D_{ij}+2k)$ where $k=0,1,\dots$ because the only cycles are length-2 cycles of the form $u \rightarrow v \rightarrow u$. The proof uses induction to prove the following inequality:    
    \begin{equation}
    \label{sec:results:tree-bound:induction hypothesis}
        N_{ij}^{(D_{ij}+2k)} \le \Delta(D_{ij}+1) (\Delta (d+2))^{k-1}, \, \forall i,j .
    \end{equation} 

    That is, the number of paths between two nodes grows exponentially in the ``excess'' length of the path over the shortest distance.
    
    \textbf{Base step 1:} for $k=1$, we have the following:
    \begin{equation}
        N_{ij}^{(D_{ij}+2)} \le \Delta (D_{ij}+1) - D_{ij} \le \Delta (D_{ij}+1),
    \end{equation}
    where the quantity in the middle is obtained by counting the number of ways of adding a length-2 cycle ($u\rightarrow v \rightarrow u$) to the shortest path from $i$ to $j$, which is bounded from above by the scenario where all nodes along the path have degree $\Delta$. Note that the boundary case of $D_{ii}=0$ is covered.
    
    \textbf{Induction 1:} suppose for some $m\ge 2$, inequality~\ref{sec:results:tree-bound:induction hypothesis} holds for all $1\le k \le (m-1)$, we want to show that it holds for $k=m$ too.

    \textbf{Base step 2:} In the case of $i=j$, any path of length $2m$ must step to a neighbor $c$ of $i$, execute a round trip of length $2m-2$, then return to $i$. Thus, we have the following:
    \begin{align}
        N_{ii}^{(D_{ii}+2m)} &= \sum_{c: c\sim i}N_{cc}^{(D_{cc}+2m-2)} \\
        &\le \Delta \Delta (D_{cc}+1) (\Delta (d+2))^{m-2} \\
        &< \Delta(D_{ii}+1) (\Delta (d+2))^{m-1},
    \end{align}
    where $c \sim i$ means that $c$ is a neighbor of $i$. The second line makes use of the fact that $i$ has $\le \Delta$ neighbors and that 
    (\ref{sec:results:tree-bound:induction hypothesis}) has been established for $k=m-1$. The third line uses $D_{cc} = D_{ii} = 0$ and $D_{cc} +1 < d+2$.
    
    \textbf{Induction 2:} Suppose (\ref{sec:results:tree-bound:induction hypothesis}) with $k=m$ has already been established for specific nodes $i$ and $j=p$. Let $c$ be a neighbor of $p$ that is not on the direct path from $i$ to $p$; in other words $p$ is the parent of $c$ as viewed from node $i$. Below, we want to show that (\ref{sec:results:tree-bound:induction hypothesis}) holds also for $j=c$:
    \begin{align}
        N_{ic}^{(D_{ic}+2m)} &= \sum_{d: d \sim c}N_{id}^{(D_{ic}+2m-1)}\\
        &= N_{ip}^{(D_{ip}+2m)} + \sum_{\substack{d: d \sim c,\\d \ne p}} N_{id}^{(D_{id}+2m-2)}\\
        &\le \Delta (D_{ip}+1) (\Delta (d+2))^{m-1} + (\Delta-1) \Delta (D_{id}+1)) (\Delta (d+2))^{m-2}\\
        &= \Delta D_{ij}\Delta^{sm-s}+(\Delta-1)\Delta(D_{ij}+2)\Delta^{sm-2s}\\
        &\le \frac{\Delta D_{ij} \Delta^{sm}}{\Delta(d+2)} + \frac{\Delta^2(d+2)\Delta^{sm}}{\Delta^2(d+2)^2}\\
        &=\Delta(D_{ij}+1)\Delta^{s(m-1)}.
    \end{align}

    Hence, inequality~\ref{sec:results:tree-bound:induction hypothesis} holds for all $k \ge 1$.
    
    Note that in particular, the right-hand side of inequality~\ref{sec:results:tree-bound:induction hypothesis} can be relaxed to be independent of $D_{ij}$:
    \begin{equation}
    \label{sec:results:tree-bound:main inequality}
         N_{ij}^{(D_{ij}+2k)} \le \Delta^{sk} = (\Delta(d+2))^k, \, \forall 1 \le i, j \le n; \; k \ge 0 .
    \end{equation}
    
    Now, we are finally able to calculate an upper bound for $Y_{ij}=(1-\gamma \mathbf{A})^{-1}_{ij}$:
    \begin{align}
        Y_{ij} &= \gamma^{D_{ij}}(\sum_{m=0}^{\infty}\gamma^{2m}N_{ij}^{(D_{ij}+2m)})\\
        &\le \gamma^{D_{ij}}[\sum_{m=0}^{\infty}(\Delta(d+2)\gamma^2)^m]\\
        &= \gamma^{D_{ij}}\frac{1}{1-\Delta(d+2)\gamma^2}.
    \end{align}
    
    The geometric series converges provided that $\gamma^2\Delta(d+2) < 1$.
    
    Imposing the condition $\gamma^{D_{ij}}\frac{1}{1-\Delta(d+2)\gamma^2} \le \gamma^{D_{ij}-1}$ gives:
    \begin{equation}
        \Delta(d+2)\gamma^2+\gamma-1 \le 0,
    \end{equation}
    which gives the quadratic root form in the statement of this proposition.
\end{proof}

\section*{Acknowledgements}
MM was supported by grants from the Simons Collaboration on the Global Brain (543015) and NIH (R01 NS111477). Thanks to Joel Tropp and Chris Umans for valuable feedback.

\section*{Code availability}
Code to produce the figures is available in a public repository: \url{https://github.com/markusmeister/APSP}.

\newpage
\section*{References}
\renewcommand{\bibsection}{}\vspace{0em}
\bibliographystyle{unsrtnat}
\bibliography{references}

\end{document}